# Fractional Programming for Communication Systems—Part I: Power Control and Beamforming

Kaiming Shen, *Student Member, IEEE*, and Wei Yu, *Fellow, IEEE*

*Abstract*—**Fractional programming (FP) refers to a family of optimization problems that involve ratio term(s). This two-part paper explores the use of FP in the design and optimization of communication systems. Part I of this paper focuses on FP theory and on solving continuous problems. The main theoretical contribution is a novel quadratic transform technique for tackling the multiple-ratio concave-convex FP problem— in contrast to conventional FP techniques that mostly can only deal with the single-ratio or the max-min-ratio case. Multiple-ratio FP problems are important for the optimization of communication networks, because system-level design often involves multiple signal-to-interference-plus-noise ratio terms. This paper considers the applications of FP to solving continuous problems in communication system design, particularly for power control, beamforming, and energy efficiency maximization. These application cases illustrate that the proposed quadratic transform can greatly facilitate the optimization involving ratios by recasting the original nonconvex problem as a sequence of convex problems. This FP-based problem reformulation gives rise to an efficient iterative optimization algorithm with provable convergence to a stationary point. The paper further demonstrates close connections between the proposed FP approach and other well-known algorithms in the literature, such as the fixed-point iteration and the weighted minimum mean-square-error beamforming. The optimization of discrete problems is discussed in Part II of this paper.**

## I. Overview

OPTIMIZATION is a key aspect of communication system design [3], [4]. This two-part work explores the application of fractional programming (FP) in the design and optimization of communication systems. FP refers to a family of optimization problems containing ratio term(s). Its history can be traced back to an early paper on economic expansion [5] by von Neumann in 1937; it has since been studied extensively in broad areas in economics, management science, information theory, optics, graph theory, and computer science [6]–[8]. For example, FP has recently been applied in [9]–[12] to solve the energy efficiency maximization problem for wireless communication systems.

Manuscript submitted to IEEE Transactions on Signal Processing January 9, 2017; revised August 31 and December 9, 2017; accepted February 13, 2018. This work is supported in part by Natural Science and Engineering Research Council (NSERC) and in part by Huawei Technologies Canada. The materials in this paper have been presented in part in Asilomar Conference on Signals, Systems, and Computers, Nov. 2015, Pacific Grove, CA [1] and in IEEE International Conference on Acoustic, Speech, and Signal Processing (ICASSP), May 2016, Shanghai, China [2]. The authors are with the Edward S. Rogers Sr. Department of Electrical and Computer Engineering, University of Toronto, Toronto, ON M5S 3G4, Canada (e-mails: {kshen,weiyu}@ece.utoronto.ca).

The aim of this two-part paper is to extend the use of FP to address a broader range of optimization problems in communication system design, in particular on power control, beamforming, and user scheduling, which often cannot be directly expressed in ratio forms. We focus on communication systems in which the data rate is computed as $\log(1 + \text{SINR})$, where SINR is the signal-to-interference-plus-noise ratio. The prominent role played by "SINR" in communication systems makes FP an invaluable tool for network design and optimization. The discussion throughout the paper focuses on wireless cellular networks, but it can be readily adapted to many other networks (e.g., the optical network or the digital subscriber lines).

Although an extensive literature already exists for FP, most of them specialize in *single-ratio* problems. For example, prior works on communication system design [9]–[12] that rely on classical FP techniques have had to limit the system model to the scenario involving only one single ratio. Although multiple-ratio problems are dealt with in [13], they are limited to specific forms (e.g., the max-min problem). System-level communication network design, however, often has to deal with multiple ratios, because the overall system performance is typically a function of multiple fractional parameters (e.g., SINRs) from multiple interfering links. Solving *multiple-ratio* FP is, however, NP-hard [14]. The state-of-the-art methods for finding globally optimal solution all require exponential running time (e.g., using branch-and-bound search [15]–[17]). In fact, as pointed out in [16] and [18], the solution to a general FP problem consisting of more than 20 ratio terms is already beyond the reach of known approaches within reasonable time. As to finding stationary-point solution of the multiple-ratio problem, only general-purpose techniques such as successive convex approximation are known.

This paper addresses the multiple-ratio FP problem from a new perspective. Our main theoretic contribution is a novel technique called *quadratic transform* that introduces some suitable auxiliary variables, then recasts the original problem to a form amenable to iterative optimization. Specifically, this new technique decouples the numerator and the denominator of each ratio term, similar to the classical *Dinkelbach's transform* (but works with multiple ratio as opposed to mostly single ratio for the classic method). This decoupling feature of the proposed quadratic transform is particularly suited for the coordinated resource optimization problem across multiple cells in a wireless cellular network. For instance, the intercell power spectrum optimization problem is a challenging nonconvex problem, because the transmit power levels of the different links strongly impact each other through the interference terms in SINR. Our proposed FP approach decouples the signal and the interference terms of the multiple links, thereby converting the original nonconvex problem into a sequence of convex problems, through a set of auxiliary variables.

Part I of this paper applies the proposed technique to solve continuous problems in communication system design. The discrete case is more challenging and is dealt with in Part II of this paper [19]. The main contributions of Part I are as follows:

- *FP Theory:* A novel technique called quadratic transform is proposed to tackle the multiple-ratio FP problems. It decouples the numerator and the denominator of each ratio term, thereby converting a concave-convex multiple-ratio FP problems into a sequence of convex optimization problems.

- *Power Control:* The proposed approach is applied to the optimization of transmit power levels that maximize the weighted sum rate of a single-input single-output (SISO) wireless cellular network, which is a challenging nonconvex problem. We propose two methods: A direct approach that applies the quadratic transform directly to SINR, then subsequently updates the power variables iteratively via a sequence of convex optimizations; and a second method that results in closed-form update in optimization process. We show the connection of the latter approach to fixed-point iteration in optimization.

- *Beamforming:* The quadratic transform is generalized to the vector case and applied to the transmit beamforming optimization problem that seeks to maximize the weighted sum rate of a multiple-input multiple-out (MIMO) wireless cellular network.

- *Energy Efficiency:* The proposed approach is applied to the maximization of the overall energy efficiency of a communication network (i.e., the ratio of the sum data rate to the total power consumption). The application of FP is ideally suited for this problem scenario, because the objective function is already in a ratio form. Prior works [9]–[12] use Dinkelbach's transform to decouple the single-ratio objective. This paper proposes a novel idea of treating the numerator itself as an inner multiple-ratio problem nested in the outer single-ratio energy efficiency problem. The resulting algorithm involves a nested use of FP.

Throughout this paper, the bold lower-case letter denotes a vector; the bold upper-case letter denotes a matrix; the calligraphy upper-case letter denotes a set. For a vector $\mathbf{a}$, $\|\mathbf{a}\|_2$ refers to its Euclidean norm; $\mathbf{a}^\dagger$ refers to its conjugate transpose. For a matrix $\mathbf{A}$, $\mathbf{A}^{-1}$ refers to its inverse, $\mathbf{A}^\dagger$ refers to its conjugate transpose. Denote $\mathbf{0}$ as the null vector, and $\mathbf{I}$ as the identity matrix. Denote $\mathbb{N}$ as the set of strictly positive integers. Define $\mathbb{R}$ as the set of real numbers, and $\mathbb{R}_+$ or $\mathbb{R}_{++}$ as the set of nonnegative or strictly positive real numbers. Denote $\mathbb{C}$ as the set of complex numbers, and $\mathrm{Re}\{\cdot\}$ as the real part. Denote $\mathbb{S}_{++}$ as the set of symmetric positive definite matrices.

## II. FRACTIONAL PROGRAMMING

FP is a class of optimization problems involving fractional terms (or ratios). This section reviews classic techniques for FP that deal with single-ratio problems, then introduces a novel quadratic transform technique capable of dealing with multiple-ratio problems.

### A. Classic Techniques

We begin by considering the single-ratio FP problem. Given a nonempty constraint set $\mathcal{X} \subseteq \mathbb{R}^d$, a nonnegative function $A(\mathbf{x})$: $\mathbb{R}^d \to \mathbb{R}_+$, and a positive function $B(\mathbf{x})$: $\mathbb{R}^d \to \mathbb{R}_{++}$, where $d \in \mathbb{N}$, a *single-ratio* (maximizing) FP problem is defined to be

$$\underset{\mathbf{x}}{\text{maximize}} \quad \frac{A(\mathbf{x})}{B(\mathbf{x})} \tag{1a}$$

$$\text{subject to} \quad \mathbf{x} \in \mathcal{X}. \tag{1b}$$

The above single-ratio FP problem is in general not convex.

The conventional approach for dealing with FP is to reformulate the problem in a form with its numerator and denominator decoupled, whereby the joint optimization of $A(\mathbf{x})$ and $B(\mathbf{x})$ becomes easier, especially for the case where $A(\mathbf{x})$ is a concave function, $B(\mathbf{x})$ is a convex function, and $\mathcal{X}$ is a convex set expressed in standard form—known as the *concave-convex* FP problem. The two classic techniques presented below relate to this type of approach.

*1) Charnes-Cooper Transform:* This classic technique for FP is proposed by the early works [20], [21]. It introduces two new variables

$$z = \frac{1}{B(\mathbf{x})} \tag{2}$$

and

$$\mathbf{q} = \frac{\mathbf{x}}{B(\mathbf{x})}, \tag{3}$$

then reformulates the single-ratio problem (1) as

$$\underset{z,\,\mathbf{q}}{\text{maximize}} \quad zA\left(\frac{\mathbf{q}}{z}\right) \tag{4a}$$

$$\text{subject to} \quad zB\left(\frac{\mathbf{q}}{z}\right) \leq 1 \tag{4b}$$

$$z \in \mathcal{Z} \tag{4c}$$

$$\mathbf{q} \in \mathcal{Q} \tag{4d}$$

where $\mathcal{Z}$ and $\mathcal{Q}$ are the range of $z$ and $\mathbf{q}$ according to (2) and (3), respectively, as $\mathbf{x} \in \mathcal{X}$. After solving the above problem, the primal solution $\mathbf{x}$ can be recovered by either (2) or (3). This transform is first proposed by Charnes and Cooper [20] for the affine case then extended by Schaible [21] to the general concave-convex case.

Observe that this technique decouples the denominator and the numerator by moving the $B(\mathbf{x})$ to the constraint (4b) while leaving $A(\mathbf{x})$ in the objective. If the problem is a concave-convex FP, then the reformulation (4) is a convex problem which can be efficiently solved. Note that in the Charnes-Cooper transform: (i) additional constraints are introduced; (ii) $\mathcal{Z}$ and $\mathcal{Q}$ need to be characterized, which may not be easy to do. We also remark that although the technique works very well for the single-ratio case (in fact converges to a global optimum solution of the concave-convex single-ratio FP problem), it cannot be easily extended to the multiple-ratio case, e.g., sum-of-ratios problem.

*2) Dinkelbach's Transform:* This classic technique, first proposed in [22], reformulates the single-ratio problem (1) as

$$\text{maximize} \quad A(\mathbf{x}) - yB(\mathbf{x}) \tag{5a}$$
$$\text{subject to} \quad \mathbf{x} \in \mathcal{X} \tag{5b}$$

with a new auxiliary variable $y$, which is iteratively updated by

$$y[t+1] = \frac{A(\mathbf{x}[t])}{B(\mathbf{x}[t])} \tag{6}$$

where $t$ is the iteration index. It can be proved that convergence is guaranteed by alternatively updating $y$ according to (6) and solving for $\mathbf{x}$ in (5), because $y$ is nondecreasing after each iteration. Specially, when the single-ratio problem (1) is a concave-convex FP, optimizing $\mathbf{x}$ in (5) for fixed $y$ is a convex problem; the overall iterative algorithm in fact converges to the global optimum solution of (1). Dinkelbach's transform has an advantage as compared to Charnes-Cooper transform in the sense that no extra constraints are introduced.

### B. Proposed Quadratic Transform

Classic transforms for FP work well for single-ratio problems, but they cannot be easily generalized to multiple-ratio FP. This is because although these classic transforms have the property that the original FP and the transformed problem have the same optimal solution, the optimal value of the objective function of the transformed problem is not necessarily the same as the original FP objective function value. Thus, when multiple ratios are involved, one cannot apply the transform to each ratio individually.

This paper proposes a new transform, which is motivated by Dinkelbach's transform, but with an added constraint that the value of the objective function must stay the same. It is named *quadratic transform* because it involves quadratic terms.

First, we formally state the properties that the desired transformed objective function must have, when reformulating the original FP objective function in (1):

C1: *(Decoupling)* The new objective has the form $g(\mathbf{x}, y) = f(A(\mathbf{x}))q_1(y) + h(B(\mathbf{x}))q_2(y)$, where $y$ is an auxiliary variable.

C2: *(Equivalent Solution)* Variable $\mathbf{x}^\star$ maximizes $A(\mathbf{x})/B(\mathbf{x})$ if and only if $\mathbf{x}^\star$ together with some $y^\star$ maximizes $g(\mathbf{x}, y)$.

C3: *(Equivalent Objective)* Let $y^\star = \arg\max_y g(\mathbf{x}, y)$ for some $\mathbf{x}$, then $g(\mathbf{x}, y^\star) = A(\mathbf{x})/B(\mathbf{x})$ for this $\mathbf{x}$.

C4: *(Concavity)* Function $g(\mathbf{x}, y)$ is concave over $y$ for fixed $\mathbf{x}$, i.e., $\partial^2 g/\partial y^2 \le 0$.

The above four conditions are all naturally motivated. C1 and C2 follow from the idea of the classic FP transforms in order to decouple the optimization of $A(\mathbf{x})$ and $B(\mathbf{x})$ through $y$; C3 makes a stronger equivalence with the original problem as motivated by the desired application for multiple-ratio problems; C4 allows for convex optimization over $y$ for fixed $\mathbf{x}$. Note that C3 implies C2 but not vice versa. In fact, Dinkelbach's transform satisfies C1, C2 and C4, but does not satisfy C3. (Specifically, at the optimum, Dinkelbach's transform has $y^\star = A(\mathbf{x})/B(\mathbf{x})$ according to (6), therefore its $g(\mathbf{x}, y^\star) = 0$.)

This paper proposes a novel *quadratic transform* for FP problem that meets all these conditions C1-C4, as stated in the following theorem.

*Theorem 1 (Quadratic Transform):* The quadratic transform

$$g(\mathbf{x}, y) = 2y\sqrt{A(\mathbf{x})} - y^2 B(\mathbf{x}) \tag{7}$$

satisfies conditions C1-C4. Further, if C4 is strengthened to require that $\partial^2 g/\partial y^2$ is independent of $y$, then any $g(\mathbf{x}, y)$ that satisfies C1-C4 must be of the form

$$g(\mathbf{x}, y) = 2(t_1 y + t_2)\sqrt{A(\mathbf{x})} - (t_1 y + t_2)^2 B(\mathbf{x}) \tag{8}$$

for some $t_1 \ne 0$ and some $t_2 \in \mathbb{R}$. Thus, the proposed quadratic transform is without loss of generality up to an affine transformation in $y$.

*Proof:* See Appendix A. ∎

### C. Quadratic Transform for Multiple-Ratio FP

We now apply the quadratic transform to general multiple-ratio FP problems. Introduce $M$ pairs of numerator functions $A_m(\mathbf{x})\colon \mathbb{R}^d \to \mathbb{R}_+$ and denominator functions $B_m(\mathbf{x})\colon \mathbb{R}^d \to \mathbb{R}_{++}$ for $m = 1, \cdots, M$, the *sum-of-ratio* problem is defined to be of the form:

$$\text{maximize} \quad \sum_{m=1}^{M} \frac{A_m(\mathbf{x})}{B_m(\mathbf{x})} \tag{9a}$$
$$\text{subject to} \quad \mathbf{x} \in \mathcal{X}. \tag{9b}$$

Condition C3 is critical for extending the idea of decoupled optimization of numerators and denominators to the sum-of-ratio problem. As mentioned before, Dinkelbach's transform does not satisfy C3. Without the equivalence in the optimal objective function value, it is normally difficult to extend Dinkelbach's transform to the multiple-ratio case (except in special cases such as the max-min problem [13]). A straight-forward extension of Dinkelbach's transform such as

$$\text{maximize} \quad \sum_{m=1}^{M} (A_m(\mathbf{x}) - y_m B_m(\mathbf{x})) \tag{10a}$$
$$\text{subject to} \quad \mathbf{x} \in \mathcal{X} \tag{10b}$$

(where $y_m$ is iteratively updated to $A_m/B_m$) does not guarantee the equivalence to (9).

In contrast, the quadratic transform in Theorem 1 can be readily extended for the sum-of-ratio problem due to C3 as shown below.

*Corollary 1:* The *sum-of-ratios* problem (9) is equivalent to

$$\text{maximize} \quad \sum_{m=1}^{M} \left(2y_m\sqrt{A_m(\mathbf{x})} - y_m^2 B_m(\mathbf{x})\right) \tag{11a}$$
$$\text{subject to} \quad \mathbf{x} \in \mathcal{X}, \ y_m \in \mathbb{R} \tag{11b}$$

where $\mathbf{y}$ refers to a collection of variables $\{y_1, \cdots, y_M\}$.

In fact, the quadratic transform can be further extended to a more general sum-of-functions-of-ratio problem and also a max-min-ratio problem, as specified in the following.

*Corollary 2:* Given a sequence of nondecreasing functions $f_m(\cdot)$ and a sequence of ratios $A_m/B_m$ for $m = 1, \ldots, M$, the *sum-of-functions-of-ratio* problem

$$\underset{\mathbf{x}}{\text{maximize}} \quad \sum_{m=1}^{M} f_m\left(\frac{A_m(\mathbf{x})}{B_m(\mathbf{x})}\right) \tag{12a}$$

$$\text{subject to} \quad \mathbf{x} \in \mathcal{X} \tag{12b}$$

is equivalent to

$$\underset{\mathbf{x}, \mathbf{y}}{\text{maximize}} \quad \sum_{m=1}^{M} f_m\left(2y_m\sqrt{A_m(\mathbf{x})} - y_m^2 B_m(\mathbf{x})\right) \tag{13a}$$

$$\text{subject to} \quad \mathbf{x} \in \mathcal{X}, \ y_m \in \mathbb{R}, \ m = 1, \ldots, M. \tag{13b}$$

To verify the corollary, we first simply rewrite problem (12) as $\max_{\mathbf{x}, \mathbf{r}} \sum_{m=1}^{M} f_m(r_m)$ subject to $\mathbf{x} \in \mathcal{X}$ and $r_m = A_m(\mathbf{x})/B_m(\mathbf{x})$; then, because of condition C3, variable $r_m$ can be replaced with $\max_y(2y_m\sqrt{A_m(\mathbf{x})} - y_m^2 B_m(\mathbf{x}))$; further, since $f_m$ is a nondecreasing function, $\max_{\mathbf{x}} \sum_{m=1}^{M} f_m(\max_{y_m}(2y_m\sqrt{A_m(\mathbf{x})} - y_m^2 B_m(\mathbf{x})))$ can be rewritten as in (13a) by combining $\max_{\mathbf{x}}$ and $\max_{\mathbf{y}}$.

*Corollary 3:* Given a sequence of ratios $A_m/B_m$ for $m = 1, \ldots, M$, the *max-min-ratio* problem

$$\underset{\mathbf{x}}{\text{maximize}} \quad \min_m\left\{\frac{A_m(\mathbf{x})}{B_m(\mathbf{x})}\right\} \tag{14a}$$

$$\text{subject to} \quad \mathbf{x} \in \mathcal{X} \tag{14b}$$

is equivalent to

$$\underset{\mathbf{x}, \mathbf{y}, z}{\text{maximize}} \quad z \tag{15a}$$

$$\text{subject to} \quad \mathbf{x} \in \mathcal{X}, \ y_m \in \mathbb{R}, \ z \in \mathbb{R}, \tag{15b}$$

$$2y_m\sqrt{A_m(\mathbf{x})} - y_m^2 B_m(\mathbf{x}) \geq z, \ \forall m. \tag{15c}$$

To verify this, we first rewrite problem (14) as $\max_{\mathbf{x}, z} z$ subject to $\mathbf{x} \in \mathcal{X}$ and $z \leq A_m(\mathbf{x})/B_m(\mathbf{x})$; because of C3, the latter constraint can be rewritten as $z \leq \max_{y_m}(2y_m\sqrt{A_m(\mathbf{x})} - y_m^2 B_m(\mathbf{x}))$; since this new constraint is a less-than-max inequality, $\max_{y_m}$ can be integrated into $\max_{\mathbf{x}, \mathbf{y}}$, as in (15).

Note that the *equivalent objective* condition C3 plays a key role in deriving the above corollaries.

### D. Multidimensional and Complex FP

We further consider FP in a multidimensional complex case where the numerators are vectors and the denominators are matrices. This class of FP arises in dealing with multi-antenna communication systems. Given a sequence of function $\mathbf{a}_m(\mathbf{x})$: $\mathbb{C}^{d_1} \to \mathbb{C}^{d_2}$ and function $\mathbf{B}_m(\mathbf{x})$: $\mathbb{C}^{d_1} \to \mathbb{S}_{++}^{d_2 \times d_2}$, for $m = 1, \ldots, M$, and constraint set $\mathcal{X} \subseteq \mathbb{C}^{d_1}$, where $d_1, d_2 \in \mathbb{N}$, a *multidimensional single-ratio* FP problem is defined to be

$$\underset{\mathbf{x}}{\text{maximize}} \quad \sum_{m=1}^{M} \mathbf{a}_m^\dagger(\mathbf{x})\mathbf{B}_m^{-1}(\mathbf{x})\mathbf{a}_m(\mathbf{x}) \tag{16a}$$

$$\text{subject to} \quad \mathbf{x} \in \mathcal{X}. \tag{16b}$$

The corresponding quadratic transform for this multidimensional case is stated in the theorem below.

*Theorem 2 (Quadratic Transform in the Multidimensional and Complex Case):* Problem (16) is equivalent to

$$\underset{\mathbf{x}, \mathbf{y}}{\text{maximize}} \quad \sum_{m=1}^{M} \left(2\text{Re}\left\{\mathbf{y}_m^\dagger \mathbf{a}_m(\mathbf{x})\right\} - \mathbf{y}_m^\dagger \mathbf{B}_m(\mathbf{x})\mathbf{y}_m\right) \tag{17a}$$

$$\text{subject to} \quad \mathbf{x} \in \mathcal{X}, \ \mathbf{y}_m \in \mathbb{C}^{d_2} \tag{17b}$$

where $\mathbf{y}$ refers to a collection of auxiliary variables $\{y_1, \cdots, y_M\}$.

*Proof:* Recognize each term in the summation of (17a) as $\mathbf{y}_m^\dagger \mathbf{a}_m + \mathbf{a}_m^\dagger \mathbf{y}_m - \mathbf{y}_m^\dagger \mathbf{B}_m \mathbf{y}_m$ and then further rewrite it as $\mathbf{a}_m^\dagger \mathbf{B}_m^{-1}\mathbf{a}_m - (\mathbf{y}_m - \mathbf{B}_m^{-1}\mathbf{a}_m^\dagger)\mathbf{B}_m(\mathbf{y}_m - \mathbf{B}_m^{-1}\mathbf{a}_m)$ by completing the square. It is easy to see that the optimal solution of (17) is $\mathbf{y}_m^\star = \mathbf{B}_m^{-1}(\mathbf{x})\mathbf{a}_m(\mathbf{x})$ and the optimal value of (17a) equals to $\mathbf{a}_m^\dagger \mathbf{B}_m^{-1}\mathbf{a}_m$ exactly. The equivalence to (16) is therefore established. ∎

This multidimensional and complex quadratic transform can be readily extended for more general sum-of-functions-of-ratio and max-min-ratio as in Corollaries 2 and 3.

### E. Iterative Optimization for Concave-Convex FP

The discussion on FP so far assumes arbitrary ratio functions (with the numerator being non-negative and the denominator being positive). We now focus on the special type of concave-convex FP problems, which is of particular importance in communication system design.

An FP problem is called *concave-convex* if it satisfies the following three conditions:

- The numerators $A_m(\mathbf{x})$ are all concave functions;
- The denominators $B_m(\mathbf{x})$ are all convex functions;
- The constraint set $\mathcal{X}$ is a nonempty convex set in standard form as expressed by a finite number of inequality constraints.

Note that a concave-convex FP problem is not necessarily a convex problem, so solving it directly can be difficult. But in some particular cases, e.g., when the problem contains only one ratio, the concave-convex FP has the desirable convexity structure that allows it to be solved globally. In fact, the aforementioned classic techniques, i.e., Charnes-Cooper transform and Dinkelbach's transform, are both initially proposed to solve the *single-ratio concave-convex* FP problem.

The main goal here is to tackle the *multiple-ratio concave-convex* FP problem using the quadratic transform. For ease of notation, we only consider the scalar case of $A_m$ and $B_m$ in what follows, but the multidimensional extension can be readily obtained according to Theorem 2.

Consider the sum-of-ratios problem (9), the sum-of-functions-of-ratio problem (12), and the max-min-ratio problem (14), but additionally assume that each $A_m(\mathbf{x})$ is concave and each $B_m(\mathbf{x})$ is convex and also that $\mathcal{X}$ is convex in standard form. Further, for the case of functions-of-ratio, we assume that the functions $f_m(\cdot)$ are not only nondecreasing, but also concave. We propose to apply the quadratic transform and optimize the primal variable $\mathbf{x}$ and the auxiliary variable $y_m$ iteratively.

---

**Algorithm 1: Iterative Approach for Concave-Convex FP Problems (9), (12), and (14)**

**Step 0:** Initialize $\mathbf{x}$ to a feasible value.

**Step 1:** Reformulate the problem by the quadratic transform, i.e., replace each $A_m/B_m$ with $2y_m\sqrt{A_m} - y_m^2 B_m$.

**Repeat**

    **Step 2:** Update $\mathbf{y}$ by (18).

    **Step 3:** Update $\mathbf{x}$ by solving the reformulated convex optimization problem (11), (13), and (15), respectively, over $\mathbf{x}$ for fixed $\mathbf{y}$.

**until** convergence.

---

When $\mathbf{x}$ is held fixed, the optimal $y_m$ can be found in closed form as

$$y_m^\star = \frac{\sqrt{A_m(\mathbf{x})}}{B_m(\mathbf{x})}, \ \forall m = 1, \ldots, M. \tag{18}$$

When $y_m$ is fixed, due to the concavity of each $A_m(\mathbf{x})$, the convexity of each $B_m(\mathbf{x})$, and that the square-root function is concave and increasing, the quadratic transform

$$g(\mathbf{x}, y_m) = 2y_m\sqrt{A_m(\mathbf{x})} - y_m^2 B_m(\mathbf{x}) \tag{19}$$

is concave in $\mathbf{x}$ for fixed $y_m$. Further, if $f_m(\cdot)$ is assumed to be concave and nondecreasing, then we also have that $f_m(g(\mathbf{x}, y_m))$ is concave in $\mathbf{x}$. Therefore, the quadratic transformed problems (11), (13) and (15) are all concave maximization problems over $\mathbf{x}$. The optimal $\mathbf{x}$ can thus be efficiently obtained through numerical convex optimization. The entire approach is summarized in Algorithm 1.

We show in the following that Algorithm 1 is guaranteed to achieve a stationary point of concave-convex FP problems.

*Theorem 3:* For the concave-convex sum-of-functions-of-ratio problem (12), i.e., every $A_m(\mathbf{x})$ is concave and every $B_m(\mathbf{x})$ is convex, and $\mathcal{X}$ is a convex set in standard form, and assuming further that $f_m$ is nondecreasing and concave, then Algorithm 1 consists of a sequence of convex optimization problems that converge to a stationary point of (12) with nondecreasing sum-of-functions-of-ratio value after every iteration.

*Proof:* The algorithm is essentially a block coordinate ascent algorithm for the reformulated problem (13), which is a convex optimization problem due to the concave-convex form of (12), so it converges to a stationary point $(\mathbf{x}^\star, \mathbf{y}^\star)$ of (13). Due to the equivalence in the solution (i.e., Condition C2) and the equivalence in the objective value (i.e., Condition C3), the first-order condition on $\mathbf{x}$ for (13) under the optimal $\mathbf{y}^\star$ is the same as for the original problem (12), hence the algorithm also converges to a stationary point of (12). Condition C3 guarantees that the sum-of-functions-of-ratio value is nondecreasing after every update of $\mathbf{y}$. ∎

Note that as the sum-of-ratios is a special case of the sum-of-functions-of-ratio, Algorithm 1 also converges to a local optimum when applied to the sum-of-ratio problem (9). For the single-ratio or max-min-ratio case, a much stronger result is possible.

*Theorem 4:* For the single-ratio problem (1) and the max-min-ratio (14) concave-convex FP problem with differentiable $A(\mathbf{x})$ and $B(\mathbf{x})$, Algorithm 1 converges to the globally optimal solution of the respective problems.

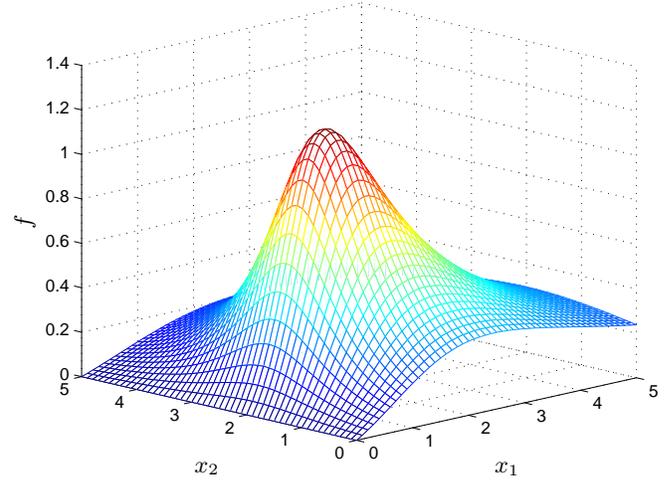

Fig. 1: Maximizing $f(x_1, x_2) = x_1/((x_1-1)^2 + (x_2-2)^2 + 1)$ over $x_1 \geq 0$ and $x_2 \geq 0$ is a single-ratio concave-convex FP problem. Although $f(x_1, x_2)$ is not concave, its stationary point is also the global optimum.

*Proof:* The key is to verify that any stationary point must be the global optimum in the special cases of single-ratio or max-min problems. This can be established by showing that the concave-convex single-ratio FP problem is *pseudo-convex*. This fact has been proved in [23] for the case where $A(\mathbf{x})$ and $B(\mathbf{x})$ are differentiable and $A(\mathbf{x})$ is concave and $B(\mathbf{x})$ is convex. Thus for single-ratio FP, Algorithm 1 converges to a global optimum. Furthermore, by the result in [24] that any local optimum solution is also the global optimum solution for the problem $\min \max_m\{f_m\}$ given that each $f_m$ is a pseudo-convex function, the global convergence of Algorithm 1 in the max-min problem case can also be established. ∎

Fig. 1 shows an example of a single-ratio concave-convex FP problem whose unique stationary point is the global optimum. We note that this property of converging to the globally optimal solution holds also for the Charnes-Cooper Transform and the Dinkelbach's transform. This is true despite that the original problem is not necessarily convex.

Algorithm 1 can be readily extended to the multidimensional and complex problem (16), i.e., by optimizing $\mathbf{y}$ and $\mathbf{x}$ alternatively in the multidimensional quadratic transform (17). The optimal $\mathbf{y}_m$ for fixed $\mathbf{x}$ is

$$\mathbf{y}_m^\star = (\mathbf{B}_m(\mathbf{x}))^{-1}\mathbf{a}_m(\mathbf{x}), \tag{20}$$

and then solving $\mathbf{x}$ for fixed $\mathbf{y}$ is a convex optimization problem under the concave-convex condition, and for the functions-of-ratio case if $f_m(\cdot)$ is concave and nondecreasing.

### F. Convergence Rate

We analyze the convergence rate of Algorithm 1 as compared to the classic transforms. Note that if the single-ratio problem is concave-convex, solving the problem by Dinkelbach's transform amounts to a sequence of convex optimiza-tions (5) over $\mathbf{x}$ with the auxiliary variable $y$ iteratively updated by (6). It is shown in [23] that the iteration by Dinkelbach's transform converges at a superlinear rate, i.e.,

$$\lim_{t\to\infty} \frac{|y^\star - y_{t+1}|}{|y^\star - y_t|} = 0 \tag{21}$$

where subscript $t$ is the index of iteration, and $y^\star$ is the auxiliary variable value at the convergence. For ease of comparison, we evaluate the convergence of Algorithm 1 for the single-ratio problem as well. As compared to Dinkelbach's transform, the quadratic transform (i.e., Algorithm 1) can be considerably slower. The following example shows that the convergence rate of Algorithm 1 can be strictly slower than superlinear.

Consider an example of single-ratio concave-convex FP

$$\underset{x}{\text{maximize}} \quad \frac{x}{x^2+1} \tag{22a}$$

$$\text{subject to} \quad x \geq 0. \tag{22b}$$

The quadratic transform reformulates its objective as

$$g(x, y) = 2y\sqrt{x} - y^2(x^2 + 1). \tag{23}$$

Introduce subscript $t$ to denote the iteration number. When $x$ is fixed at $x_t$, the optimal $y$ is updated by (18)

$$y_{t+1} = \frac{\sqrt{x_t}}{x_t^2+1}. \tag{24}$$

After $y$ is updated to $y_{t+1}$, the optimal $x$ is found to be (by solving the convex problem analytically)

$$x_{t+1} = (2y_{t+1})^{-\frac{2}{3}}. \tag{25}$$

These two updates amount to

$$y_{t+1} = \frac{(2y_t)^{-\frac{1}{3}}}{(2y_t)^{-\frac{4}{3}} + 1}. \tag{26}$$

With $y$ initialized to 0.1 (i.e., $y_0 = 0.1$), it can be shown that $y_{t+1}$ in (26) converges to $\frac{1}{2}$ in a nondecreasing fashion. We then have

$$\lim_{t\to\infty} \frac{|y^\star - y_{t+1}|}{|y^\star - y_t|} = \lim_{t\to\infty} \frac{y^\star - y_{t+1}}{y^\star - y_t} \tag{27a}$$

$$= \lim_{y\to\frac{1}{2}} \frac{1}{\frac{1}{2} - y} \left( \frac{1}{2} - \frac{(2y)^{-\frac{1}{3}}}{(2y)^{-\frac{4}{3}} + 1} \right) \tag{27b}$$

$$= \frac{1}{3}. \tag{27c}$$

Thus, Algorithm 1 in this example converges more slowly than the iterative optimization based on Dinkelbach's transform. The convergence of these two methods is illustrated in Fig. 2.

We emphasize that although the conventional Dinkelbach's transform can result in a faster convergence rate than the proposed quadratic transform, the use of the former technique is restricted to the single-ratio problem whereas the latter is capable of dealing with multiple ratios. Further, for multiple-ratio FP problems where global convergence is not guaranteed, slower convergence can sometime be advantageous as it allows the algorithm to more fully explore the solution space.

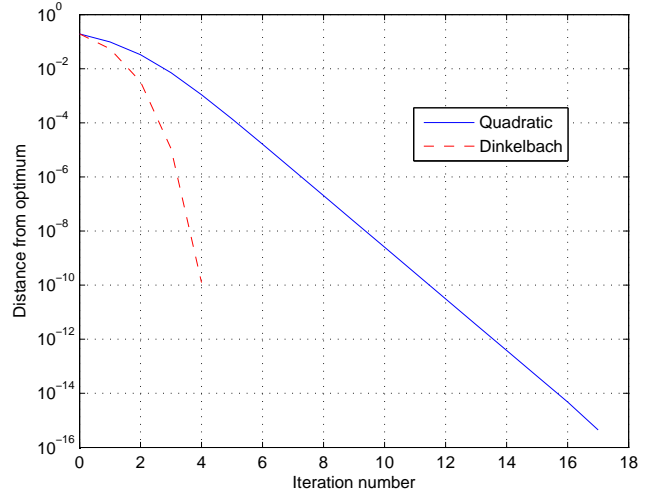

Fig. 2: When applied to the single-ratio problem (22), Dinkelbach's transform converges faster than the quadratic transform.

## III. Power Control

### A. Problem Statement

We now consider the application of FP to communication system design. The first example is the classic power control problem for a downlink SISO cellular network with a set of single-antenna base stations (BSs) $\mathcal{B}$, each serving a single-antenna user. Let $h_{i,j} \in \mathbb{C}$ be the downlink channel from BS $j$ to user $i$; let $\sigma^2$ be the additive white Gaussian noise (AWGN) power level. Introduce variable $p_i$ for each BS $i$ as its transmit power level, constrained by a power budget of $P_{\max}$. The downlink data rate of user $i$ is computed as[1]

$$R_i = \log\left(1 + \frac{|h_{i,i}|^2 p_i}{\sum_{j\neq i} |h_{i,j}|^2 p_j + \sigma^2}\right). \tag{28}$$

We consider the maximization of a weighted sum rate objective of

$$f_o(\mathbf{p}) = \sum_{i\in\mathcal{B}} w_i R_i \tag{29}$$

where $w_i$ accounts for the priority of the $i$th BS-user downlink and $\mathbf{p}$ refers to the collection $\{p_i\}_{i\in\mathcal{B}}$. The power control problem is formulated as

$$\underset{\mathbf{p}}{\text{maximize}} \quad f_o(\mathbf{p}) \tag{30a}$$

$$\text{subject to} \quad 0 \leq p_i \leq P_{\max}, \; \forall i \in \mathcal{B}. \tag{30b}$$

This problem is difficult to solve because it is nonconvex. Indeed, the problem can be solved globally by using a polyblock approximation approach [25], but not in polynomial time. Moreover, for the case where all the SINRs are sufficiently high so that $\log(1+\text{SINR})$ can be approximated as $\log(\text{SINR})$, the problem can be globally solved via *geometric programming* [4]. This paper aims to find at least a stationary point in an efficient manner. We remark that the power control problem has been studied extensively in the literature, e.g., the

---

[1]For ease of notation, we use the natural logarithm in $\log(1 + \text{SINR})$.

---

**Algorithm 2: Direct FP for Power Control**

---

**Step 0:** Initialize $\mathbf{p}$ to a feasible value.
**Repeat**
    **Step 1:** Update $\mathbf{y}$ by (33).
    **Step 2:** Update $\mathbf{p}$ by solving the convex optimization problem (31) over $\mathbf{p}$ for fixed $\mathbf{y}$.
**until** the value of function $f_q^{\mathrm{DIR}}$ in (32) converges.

---

structure of the interference functions is investigated in [26], [27].

### B. Direct FP Approach

Although the power control problem is not in a direct ratio form, the main components of its objective function, the SINR terms, are in fractional form. Because each SINR term resides inside the logarithm function, which is nondecreasing and concave, the condition of Theorem 3 is satisfied in this problem.

Specifically, after applying the quadratic transform to each SINR term, we arrive at the following reformulation

$$\underset{\mathbf{p},\,\mathbf{y}}{\text{maximize}} \qquad f_q^{\mathrm{DIR}}(\mathbf{p},\mathbf{y}) \qquad (31\mathrm{a})$$

$$\text{subject to} \qquad 0 \leq p_i \leq P_{\max},\ \forall i \in \mathcal{B} \qquad (31\mathrm{b})$$

$$y_i \in \mathbb{R},\ \forall i \in \mathcal{B} \qquad (31\mathrm{c})$$

where $\mathbf{y}$ refers to the collection $\{y_i\}_{i \in \mathcal{B}}$. The new objective $f_q^{\mathrm{DIR}}$ is

$$f_q^{\mathrm{DIR}}(\mathbf{x},\mathbf{y}) = \sum_{i \in \mathcal{B}} w_i \log\left(1 + 2y_i\sqrt{|h_{i,i}|^2 p_i} - y_i^2\left(\sum_{j \neq i} |h_{i,j}|^2 p_j + \sigma^2\right)\right). \quad (32)$$

where $y_i$ is introduced by the quadratic transform for each downlink $i$.

Following Algorithm 1, we optimize $y_i$ and $p_i$ in an iterative fashion. The optimal $y_i$ for fixed $p_i$ is

$$y_i^\star = \frac{\sqrt{|h_{i,i}|^2 p_i}}{\sum_{j \neq i} |h_{i,j}|^2 p_j + \sigma^2}. \qquad (33)$$

Then, finding the optimal $p_i$ for fixed $y_i$ is a convex problem. This power control method is summarized in Algorithm 2. By Theorem 3, Algorithm 2 guarantees a convergence to a stationary point of problem (30).

We remark that Algorithm 2 can be easily extended to the multiple-band system, where the frequency band is partitioned into $T$ sub-bands, and the user rate is computed as

$$R_i = \sum_{t=1}^{T} \frac{1}{T} \log\left(1 + \frac{|h_{i,i}^t|^2 p_i^t}{\sum_{j \neq i} |h_{i,j}^t|^2 p_j^t + \sigma^2}\right). \qquad (34)$$

Here, $h_{i,j}^t$ and $p_j^t$ represent the channel and the transmit power level in the $t$th sub-band, respectively. The power constraint

(30b) now becomes

$$\sum_{t=1}^{T} p_i^t \leq P_{\max},\ \forall i \in \mathcal{B} \qquad (35)$$

and

$$p_i^t \geq 0,\ \forall i \in \mathcal{B},\ t = 1, \ldots, T. \qquad (36)$$

To modify Algorithm 2 to work in this multiple-band scenario: Step 1 remains the same; Step 2 updates $\mathbf{p}$ by solving a convex problem.

As a final remark in this subsection, the direct FP approach for power control can be adapted to the maximization of a general rate utility function in wireless networks, as stated in the proposition below.

*Proposition 1 (General Utility Maximization):* Given a nondecreasing concave utility function $U_i$ of rate $R_i$ for each user $i$, the sum utility maximizing problem

$$\underset{\mathbf{p}}{\text{maximize}} \qquad \sum_{i \in \mathcal{B}} U_i(R_i) \qquad (37\mathrm{a})$$

$$\text{subject to} \qquad 0 \leq p_i \leq P_{\max},\ \forall i \in \mathcal{B} \qquad (37\mathrm{b})$$

is equivalent to

$$\underset{\mathbf{p},\,\mathbf{y}}{\text{maximize}} \qquad \sum_{i \in \mathcal{B}} U_i(Q_i) \qquad (38\mathrm{a})$$

$$\text{subject to} \qquad 0 \leq p_i \leq P_{\max},\ \forall i \in \mathcal{B} \qquad (38\mathrm{b})$$

$$y_i \in \mathbb{R},\ \forall i \in \mathbb{R} \qquad (38\mathrm{c})$$

where

$$Q_i = \log\left(1 + 2y_i|h_{i,i}|\sqrt{p_i} - y_i^2\sum_{j \neq i} |h_{i,j}|^2 p_j - y_i^2\sigma^2\right). \quad (39)$$

The above reformulated problem can be solved (to a stationary point) as follows. When $\mathbf{p}$ is fixed, variable $\mathbf{y}$ is optimally determined by (33); when $\mathbf{y}$ is fixed, optimizing $\mathbf{p}$ in (38) is a convex problem.

Furthermore, we remark that the direct FP approach can be also readily applied to the problem of optimizing power for maximizing the minimum rate across the users, according to Corollary 3.

### C. Closed-Form FP Approach

This section presents a different use of FP for solving the power control problem. This new approach is based on a Lagrangian dual reformulation of the power control problem as stated below. This leads to an algorithm in which each iteration is performed in closed form, rather than having to solve a convex optimization problem numerically, which is often more desirable than the direct FP approach introduced in the previous section.

*Proposition 2:* The original power control problem (30) is equivalent to

$$\underset{\mathbf{x},\,\boldsymbol{\gamma}}{\text{maximize}} \qquad f_r^{\mathrm{CF}}(\mathbf{x},\boldsymbol{\gamma}) \qquad (40\mathrm{a})$$

$$\text{subject to} \qquad \mathbf{x} \in \mathcal{X} \qquad (40\mathrm{b})$$

---

**Algorithm 3: Closed-Form FP for Power Control**

**Step 0:** Initialize $\mathbf{p}$ and $\boldsymbol{\gamma}$ feasible values.

**Repeat**

    **Step 1:** Update $\mathbf{y}$ by (45).

    **Step 2:** Update $\boldsymbol{\gamma}$ by (42).

    **Step 3:** Update $\mathbf{p}$ by (44).

**until** the value of function $f_q^{\text{CF}}$ in (41) converges.

---

where $\boldsymbol{\gamma}$ refers to a set of auxiliary variables $\{\gamma_i\}_{i\in\mathcal{B}}$, and the new objective is

$$f_r^{\text{CF}}(\mathbf{p},\boldsymbol{\gamma}) = \sum_{i\in\mathcal{B}} w_i \log\left(1+\gamma_i\right) - \sum_{i\in\mathcal{B}} w_i \gamma_i$$
$$+ \sum_{i\in\mathcal{B}} \frac{w_i(1+\gamma_i)|h_{i,i}|^2 p_i}{\sum_{j\in\mathcal{B}} |h_{i,j}|^2 p_j + \sigma^2}. \quad (41)$$

*Proof:* We defer a detailed constructive proof to Part II of the paper [19]. ∎

We propose an iterative algorithm based on the above reformulation. When $p_i$ is held fixed, the optimal $\gamma_i$ is obtained by setting $\partial f_r^{\text{CF}}/\partial \gamma_i$ to zero, i.e.,

$$\gamma_i^\star = \frac{|h_{i,i}|^2 p_i}{\sum_{j\neq i} |h_{i,j}|^2 p_j + \sigma^2}, \; \forall i \in \mathcal{B}. \quad (42)$$

Note that the optimal $\gamma_i$ is equal to the downlink SINR of BS $i$. When $\gamma_i$ is held fixed, only the last term of $f_r^{\text{CF}}$, which has a sum-of-ratio form, is involved in the optimization of $p_i$. By the quadratic transform, we further recast $f_r^{\text{CF}}$ to

$$f_q^{\text{CF}}(\mathbf{p},\boldsymbol{\gamma},\mathbf{y}) = \sum_{i\in\mathcal{B}} 2y_i\sqrt{w_i(1+\gamma_i)|h_{i,i}|^2 p_i}$$
$$- \sum_{i\in\mathcal{B}} y_i^2\left(\sum_{j\in\mathcal{B}} |h_{i,j}|^2 p_j + \sigma^2\right) + \text{const}(\boldsymbol{\gamma}) \quad (43)$$

where $\mathbf{y}$ refers to the set $\{y_i\}_{i\in\mathcal{B}}$ and $\text{const}(\boldsymbol{\gamma})$ refers to a constant term when $\boldsymbol{\gamma}$ is fixed. For maximizing $f_q^{\text{CF}}$ iteratively over $p_i$ and $y_i$, we find closed-form update equations as

$$p_i^\star = \min\left\{P_{\max}, \frac{y_i^2 w_i(1+\gamma_i)|h_{i,i}|^2}{\left(\sum_{j\in\mathcal{B}} y_j^2 |h_{j,i}|^2\right)^2}\right\}, \; \forall i \in \mathcal{B} \quad (44)$$

and

$$y_i^\star = \frac{\sqrt{w_i(1+\gamma_i)|h_{i,i}|^2 p_i}}{\sum_{j\in\mathcal{B}} |h_{i,j}|^2 p_j + \sigma^2}, \; \forall i \in \mathcal{B}. \quad (45)$$

These updating steps amount to an iterative optimization as stated in Algorithm 3. Unlike the direct FP approach, the above algorithm is not a conventional block coordinate ascent, because the optimizing objective is not fixed, i.e., $\gamma_i$ is optimally updated for $f_r^{\text{CF}}$ while $y_i$ and $p_i$ are optimally updated for $f_q^{\text{CF}}$. Nonetheless, its convergence to the stationary point can still be established. We defer the proof to Part II [19, Appendix A].

As a remark, Algorithms 2 and 3 can be initialized with simple but reasonable heuristic. For example, the initial power level $\mathbf{p}$ may be set to the half of the max power. In the simulation results in this paper, in order to guarantee fair comparisons, we use random starting points then average out the results. Moreover, we set some small constant $\delta > 0$ and use the convergence criterion $|f_q^{(t)} - f_q^{(t-1)}| < \delta$ where $t$ is the iteration index.

### D. Connection with Fixed-Point Iteration

This subsection illustrates that Algorithm 3 can be interpreted as a fixed-point iteration on the first-order condition of the power optimization problem. Attaining a stationary-point solution of the power control problem is equivalent to finding a solution to the first-order condition for (30), i.e.,

$$\frac{\partial f_o(\mathbf{p})}{\partial p_i} = 0, \; \forall i \in \mathcal{B} \quad (46)$$

which can be written as

$$\frac{1}{p_i} \cdot \underbrace{\frac{w_i \gamma_i(\mathbf{p})}{1+\gamma_i(\mathbf{p})}}_{T_{1i}(\mathbf{p})} - \underbrace{\sum_{j\neq i} \frac{w_j \gamma_j^2(\mathbf{p})|h_{j,i}|^2}{(1+\gamma_i(\mathbf{p}))|h_{j,j}|^2 p_j}}_{T_{2i}(\mathbf{p})} = 0 \quad (47)$$

where $\gamma_i(\mathbf{p})$ denotes the SINR function of $\mathbf{p}$ in cell $i$ as defined in (42). To find a set of powers that satisfy the above condition, one strategy [28]–[30] is to isolate $p_i$ at one side of the equation—this automatically results in an update equation for power, which, if converging, would achieve at least a stationary point of the power control problem.

However, it is in general not easy to decide which part of the left-hand side of (46) should be fixed in order to ensure the convergence of fixed-point iteration. For instance, [29] proposes to fix $T_{1i}$ and $T_{2i}$ as shown in (47) and arrives at the following fixed-point method for power control

$$p_i[t+1] = \min\left\{P_{\max}, \frac{T_{1i}(\mathbf{p}[t])}{T_{2i}(\mathbf{p}[t])}\right\}, \; \forall i \in \mathcal{B} \quad (48)$$

where the index $t$ indicates the iteration number. However, this fixed-point iteration does not necessarily converge. (In fact, [29] proves that this iteration is guaranteed to converge when the resulting SINR values are all sufficiently high.)

With $\boldsymbol{\gamma}^\star$ and $\mathbf{y}^\star$ substituted in (44), the update equation (44) can also be thought of as a fixed-point iteration of the first-order condition for power control, exactly like (47) except that different components $\widetilde{T}_{1i}$ and $\widetilde{T}_{2i}$, shown below, are fixed

$$\frac{1}{\sqrt{p_i}} \cdot \underbrace{\frac{w_i \gamma_i(\mathbf{p})}{\sqrt{p_i}}}_{\widetilde{T}_{1i}(\mathbf{p})} - \underbrace{\sum_j \frac{w_j \gamma_j^2(\mathbf{p})|h_{j,i}|^2}{(1+\gamma_i(\mathbf{p}))|h_{j,j}|^2 p_j}}_{\widetilde{T}_{2i}(\mathbf{p})} = 0. \quad (49)$$

In this case, the transmit power variable $p_i$ update becomes

$$p_i[t+1] = \min\left\{P_{\max}, \left(\frac{\widetilde{T}_{1i}(\mathbf{p}[t])}{\widetilde{T}_{2i}(\mathbf{p}[t])}\right)^2\right\}, \; \forall i \in \mathcal{B}, \quad (50)$$

which, along with an additional projection step onto the constraint set, can be seen to be (44) after some algebra. Thus, the power control part of Algorithm 3 is just a fixed-point iteration, but with a crucial advantage that convergence is guaranteed, in contrast to the updates proposed in [28]–[30].

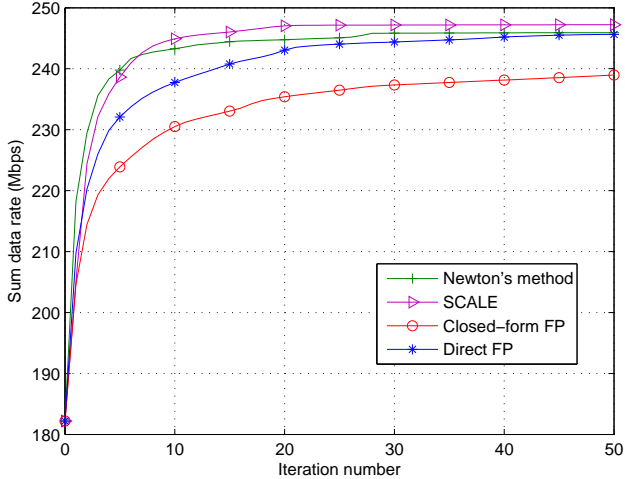

Fig. 3: Power control in flat-fading channels.

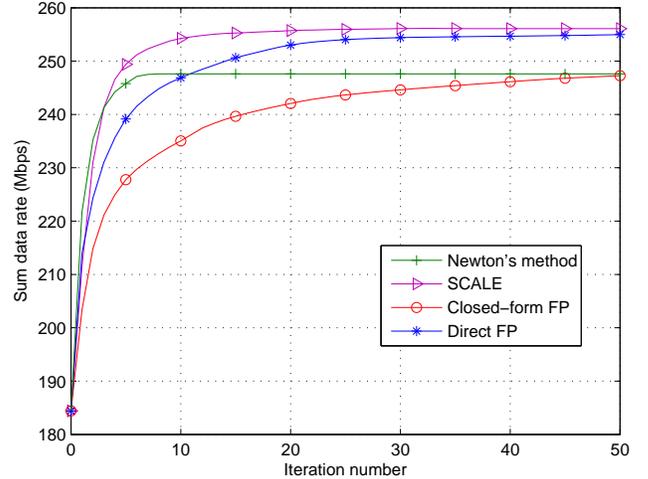

Fig. 4: Power control in frequency-selective fading channels.

### E. Numerical Example

We now evaluate the performance of FP for power control on a downlink cellular network consisting of seven wrapped-around hexagonal cells. Within each cell, the BS is located at the center and the downlink users are randomly placed. The BS-to-BS distance is set to be 0.8km. The maximum transmit power level at the BS side is set to be 43dBm, and the AWGN power level is set to be $-100$dBm. A 10MHz frequency band is fully reused across all the cells. The downlink distance-dependent path-loss is simulated by $128.1 + 37.6 \log_{10}(d) + \tau$ (in dB), where $d$ represents the BS-to-user distance in km, and $\tau$ is a zero-mean Gaussian random variable with 8dB standard deviation for the shadowing effect. We consider sum rate maximization by setting all the weights to 1.

The proposed FP approaches are compared to several benchmarks: first, direct optimization based on a modified Newton's method [31], which deals with the power constraints via the nearest-point projection (the full Newton's method is too computationally complex), and second, an approach based on a modified version of geometric programming (GP) called SCALE [32]. The version of SCALE implemented here involves solving a GP in every iteration.

Fig. 3 shows the performance of various power control algorithms in flat-fading channels. The closed-form FP takes the largest number of iterations to converge, but its computation per iteration is the lowest because of the closed-form updates in every iteration. In contrast, SCALE and direct FP both require solving a convex problem in each iteration. The closed-form FP also has lower complexity than Newton's method on per-iteration basis. In our simulation experience, the closed-form FP is the fastest.

Fig. 4 simulates a frequency selective fading scenario, in which the bandwidth is divided into 4 subbands; one downlink user is scheduled per tone. The resulting power control differs from the flat-fading case because of the sum power constraint across the subbands, i.e., $\sum_n p_i^n \le P_{\max}$ where $p_i^n$ denotes the power level in tone $n$ at BS $i$. In this case, Newton's method has to apply a heuristic nearest-point projection in order to satisfy the sum power constraint, but this no longer guarantees a stationary-point solution. As can be seen in the simulation, Newton's method now has much worse performance.

To conclude, the FP based approaches are competitive with the state-of-the-art algorithms in power control, with the closed-form FP having lower overall complexity due to its lower per-iteration cost. Note that the converged values of different algorithms may differ depending on the starting point, as only stationary-point convergence is guaranteed in all cases.

## IV. BEAMFORMING

### A. Problem Statement

The second example is an application of multidimensional FP to the beamforming optimization problem. Consider a downlink MIMO cellular network with a set of BSs $\mathcal{B}$. Assume that each BS has $M$ antennas and each user terminal has $N$ antennas; then at most $M$ downlink data streams are supported per cell via spatial multiplexing. Let $\mathbf{H}_{im,j} \in \mathbb{C}^{N \times M}$ be the downlink channel from BS $j$ to the user who is scheduled in the $m$th data stream at BS $i$. Let $\sigma^2$ be the AWGN power level. Introduce variable $\mathbf{v}_{im} \in \mathbb{C}^M$ as the downlink transmit beamformer at BS $i$ for its $m$-th data stream. The data rate of stream $(i, m)$, $R_{im}$, is computed by (51) as shown at the bottom of the page.

$$R_{im}(\mathbf{V}) = \log\left(1 + \mathbf{v}_{im}^\dagger \mathbf{H}_{im,i}^\dagger \left(\sigma^2 \mathbf{I} + \sum_{(j,n) \neq (i,m)} \mathbf{H}_{im,j} \mathbf{v}_{jn} \mathbf{v}_{jn}^\dagger \mathbf{H}_{im,j}^\dagger\right)^{-1} \mathbf{H}_{im,i} \mathbf{v}_{im}\right) \tag{51}$$

Let weight $w_{im}$ be the priority of user scheduled in the $m$-th data stream at BS $i$. We seek to maximize the weighted sum rate over the beamforming vectors:

$$\underset{\mathbf{V}}{\text{maximize}} \quad \sum_{i,m} w_{im} R_{im}(\mathbf{V}) \tag{52a}$$

$$\text{subject to} \quad \sum_{m=1}^{M} \|\mathbf{v}_{im}\|_2^2 \leq P_{\max}, \ \forall i \in \mathcal{B} \tag{52b}$$

where $\mathbf{V}$ refers to the collection $\{\mathbf{v}_{im}\}$, $P_{\max}$ refers to the transmit power budget at the BS side. This is a challenging nonconvex problem with vector variables.

### B. Multidimensional Direct FP Approach

Similar to the power control case, the direct FP approach applies the multidimensional quadratic transform (Theorem 2) to each SINR term. This leads to a new objective $f_q^{\text{DIR}}$ as in (53) at the bottom of the page, where $\mathbf{Y}$ refers to a collection of auxiliary variables $\{\mathbf{y}_{im}\}$ with $\mathbf{y}_{im} \in \mathbb{C}^N$ introduced for each data stream $(i, m)$. The optimization problem (52) can now be recast to

$$\underset{\mathbf{V},\mathbf{Y}}{\text{maximize}} \quad f_q^{\text{DIR}}(\mathbf{V}, \mathbf{Y}) \tag{54a}$$

$$\text{subject to} \quad \sum_{m=1}^{M} \|\mathbf{v}_{im}\|_2^2 \leq P_{\max}, \ \forall i \in \mathcal{B} \tag{54b}$$

$$\mathbf{y}_{im} \in \mathbb{C}^N. \tag{54c}$$

Decoupled by the multidimensional quadratic transform, the SINR term is converted to a concave function of $\mathbf{v}_{im}$. Since the outer logarithmic function is nondecreasing and concave, the optimization problem (54) is a convex problem of $\mathbf{v}_{im}$ when the auxiliary variable $\mathbf{y}_{im}$ is held fixed.

We follow Algorithm 1 to maximize $f_q^{\text{DIR}}$ over $\mathbf{v}_{im}$ and $\mathbf{y}_{im}$ iteratively. The optimal $\mathbf{y}_{im}$ for fixed $\mathbf{v}_{im}$ is

$$\mathbf{y}_{im}^\star = \left(\sigma^2\mathbf{I} + \sum_{(j,n)\neq(i,m)} \mathbf{H}_{im,j}\mathbf{v}_{jn}\mathbf{v}_{jn}^\dagger\mathbf{H}_{im,j}^\dagger\right)^{-1} \mathbf{H}_{im,i}\mathbf{v}_{im}. \tag{55}$$

---

**Algorithm 4: Direct FP for Beamforming**

**Step 0:** Initialize $\mathbf{V}$ to a feasible value.
**Repeat**
    **Step 1:** Update $\mathbf{Y}$ by (55).
    **Step 2:** Update $\mathbf{V}$ by solving the convex problem (54) over $\mathbf{V}$ for fixed $\mathbf{Y}$.
**until** the value of function $f_q^{\text{DIR}}$ in (53) converges.

---

For fixed $\mathbf{y}_{im}$, the optimal $\mathbf{v}_{im}$ can be obtained by convex optimization. The resulting algorithm, stated as Algorithm 4, has a provable convergence to a stationary point due to Theorem 3.

This algorithm requires solving a convex problem numerically in every iteration. In the next section, we illustrate another use of FP that yields a closed-form optimization in every iteration.

### C. Multidimensional Closed-Form FP Approach

As for power control, a closed-form FP approach can also be developed for the beamforming problem. The main idea is the same as in power control, but in a multidimensional vector space. The sum logarithm problem is first reformulated in a sum-of-ratio form using a Lagrangian dual transform; the quadratic transform is subsequently applied to the ratios. After applying a multidimensional extension of Proposition 2 to (52), we arrive at a sum-of-ratio reformulation with $f_r^{\text{CF}}(\mathbf{V}, \boldsymbol{\gamma})$ as in (56) at the bottom of the page, where $\boldsymbol{\gamma}$ refers to the collection $\{\gamma_{im}\}$. Again, we defer the proof of the Lagrangian dual transform to Part II of the paper [19].

When $\mathbf{v}_{im}$ is fixed, the optimal $\gamma_{im}$ can be found by setting $\partial f_r^{\text{CF}}/\partial\gamma_{im}$ to zero with respect to each $(i, m)$ tuple, as shown in (57) at the bottom of the page.

The multidimensional quadratic transform in Theorem 2 can then be readily applied to further recast $f_r^{\text{CF}}$ to $f_q^{\text{CF}}$ in (58) as displayed at the bottom of the page, where $\mathbf{Y}$ is the collection $\{y_{im}\}$ and $\text{const}(\boldsymbol{\gamma})$ is a constant term when $\boldsymbol{\gamma}$ is fixed.

---

$$f_q^{\text{DIR}}(\mathbf{V}, \mathbf{Y}) = \sum_{(i,m)} w_{im} \log\left(1 + 2\text{Re}\left\{\mathbf{y}_{im}^\dagger\mathbf{H}_{im,i}\mathbf{v}_{im}\right\} - \mathbf{y}_{im}^\dagger\left(\sigma^2\mathbf{I} + \sum_{(j,n)\neq(i,m)} \mathbf{H}_{im,j}\mathbf{v}_{jn}\mathbf{v}_{jn}^\dagger\mathbf{H}_{im,j}^\dagger\right)\mathbf{y}_{im}\right) \tag{53}$$

---

$$f_r^{\text{CF}}(\mathbf{V}, \boldsymbol{\gamma}) = \sum_{(i,m)} w_{im}\left(\log(1 + \gamma_{im}) - \gamma_{im} + (1 + \gamma_{im})\mathbf{v}_{im}^\dagger\mathbf{H}_{im,i}^\dagger\left(\sigma^2\mathbf{I} + \sum_{(j,n)} \mathbf{H}_{im,j}\mathbf{v}_{jn}\mathbf{v}_{jn}^\dagger\mathbf{H}_{im,j}^\dagger\right)^{-1}\mathbf{H}_{im,i}\mathbf{v}_{im}\right) \tag{56}$$

---

$$\gamma_{im}^\star = \mathbf{v}_{im}^\dagger\mathbf{H}_{im,i}^\dagger\left(\sigma^2\mathbf{I} + \sum_{(j,n)\neq(i,m)} \mathbf{H}_{im,j}\mathbf{v}_{jn}\mathbf{v}_{jn}^\dagger\mathbf{H}_{im,j}^\dagger\right)^{-1}\mathbf{H}_{im,i}\mathbf{v}_{im} \tag{57}$$

---

$$f_q^{\text{CF}}(\mathbf{V}, \boldsymbol{\gamma}, \mathbf{Y}) = \sum_{(i,m)}\left(2\sqrt{w_{im}(1 + \gamma_{im})}\,\text{Re}\{\mathbf{v}_{im}^\dagger\mathbf{H}_{im,i}^\dagger\mathbf{y}_{im}\} - \mathbf{y}_{im}^\dagger\left(\sigma^2\mathbf{I} + \sum_{(j,n)} \mathbf{H}_{im,j}\mathbf{v}_{jn}\mathbf{v}_{jn}^\dagger\mathbf{H}_{im,j}^\dagger\right)\mathbf{y}_{im}\right) + \text{const}(\boldsymbol{\gamma}) \tag{58}$$

---

**Algorithm 5: Closed-Form FP for Beamforming**

**Step 0:** Initialize $\mathbf{V}$ and $\boldsymbol{\gamma}$ to feasible values.

**Repeat**

    **Step 1:** Update $\mathbf{Y}$ by (60).

    **Step 2:** Update $\boldsymbol{\gamma}$ by (57).

    **Step 3:** Update $\mathbf{V}$ by (61).

**until** the value of function $f_q^{\mathrm{CF}}$ in (58) converges.

---

The above $f_q^{\mathrm{CF}}$ reformulation is obtained by treating $\sqrt{w_{im}(1+\gamma_{im})}\mathbf{H}_{im,i}\mathbf{v}_{im}$ as the numerator vector and also treating $\sigma^2\mathbf{I}+\sum_{(j,n)}\mathbf{H}_{im,j}\mathbf{v}_{jn}\mathbf{v}_{jn}^\dagger\mathbf{H}_{im,j}^\dagger$ as the denominator matrix in Theorem 2. Problem (52) is then reformulated as

$$\underset{\mathbf{V},\,\mathbf{Y}}{\text{maximize}} \qquad f_q^{\mathrm{CF}}(\mathbf{V},\boldsymbol{\gamma},\mathbf{Y}) \tag{59a}$$

$$\text{subject to} \qquad \sum_{m=1}^{M}\|\mathbf{v}_{im}\|_2^2 \le P_{\max},\ \forall i\in\mathcal{B} \tag{59b}$$

$$\gamma_{im}\in\mathbb{R},\ \mathbf{y}_{im}\in\mathbb{C}^N. \tag{59c}$$

The merit of reformulating $f_r^{\mathrm{CF}}$ as $f_q^{\mathrm{CF}}$ is to facilitate iterative optimization over $\mathbf{v}_{im}$. With the other variables fixed, the optimal $\mathbf{y}_{im}$ can be found by solving $\partial f_q^{\mathrm{CF}}/\partial\mathbf{y}_{im}=\mathbf{0}$, i.e.,

$$\mathbf{y}_{im}^\star = \left(\sigma^2\mathbf{I}+\sum_{(j,n)}\mathbf{H}_{im,j}\mathbf{v}_{jn}\mathbf{v}_{jn}^\dagger\mathbf{H}_{im,j}^\dagger\right)^{-1}\cdot\sqrt{w_{im}(1+\gamma_{im})}\mathbf{H}_{im,i}\mathbf{v}_{im}. \tag{60}$$

Likewise, the optimal $\mathbf{V}$ is

$$\mathbf{v}_{im}^\star = \left(\eta_i\mathbf{I}+\sum_{(j,n)}\mathbf{H}_{jn,i}^\dagger\mathbf{y}_{jn}\mathbf{y}_{jn}^\dagger\mathbf{H}_{jn,i}\right)^{-1}\cdot\sqrt{w_{im}(1+\gamma_{im})}\mathbf{H}_{im,i}^\dagger\mathbf{y}_{im} \tag{61}$$

where $\eta_i$ is a dual variable introduced for the power constraint, optimally determined by (due to complementary slackness)

$$\eta_i^\star = \min\left\{\eta_i\ge 0:\sum_{m=1}^{M}\|\mathbf{v}_{im}(\eta_i)\|_2^2\le P_{\max}\right\}. \tag{62}$$

Note that the optimal $\eta_i$ in (62) can be determined efficiently by bisection search. Algorithm 5 summarizes the above steps.

We remark that the proposed FP framework in this particular beamforming case, i.e., Algorithm 5, is equivalent to the well-known WMMSE algorithm [33], [34]. (This can be verified by substituting $\boldsymbol{\gamma}$ and $\mathbf{Y}$ in the updating formula of $\mathbf{V}$). We explore this connection further in Part II of the paper [19]. Like Algorithm 3, Algorithm 5 is not a block coordinate ascent but its convergence can be established. The proof is deferred to Part II [19, Appendix A].

### D. Numerical Example

The simulation model assumes the same setting as in Section IV-D for network topology, AWGN, distance-dependent pathloss, max transmit power, except that two users are randomly located within each cell and that the BSs and the users are now equipped with 2 antennas each. Consider Rayleigh fading for the channel coefficients. We pursue a maximization of sum rate in the network by setting all the weights $w_{im}=1$.

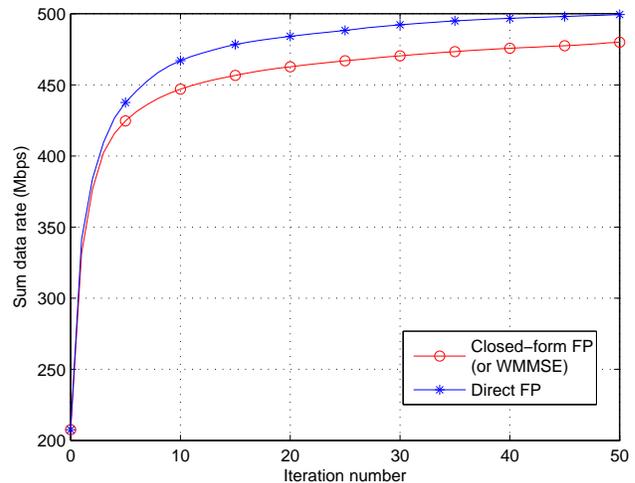

Fig. 5: Beamforming for sum data rate maximization.

Fig. 5 compares the different FP approaches. It shows that direct FP converges in fewer iterations than the closed-form FP, e.g., the former achieves a sum rate of 470Mbps within 10 iterations but the latter needs 25 iterations. However, counting just the number of iterations is misleading. The closed-form FP is in fact much more efficient than direct FP on a per-iteration basis, because closed-form FP updates all variables in closed form, while direct FP requires solving a convex optimization in each iteration. Therefore, the closed-form FP algorithm is much preferred.

## V. Energy Efficiency Maximization

As a final example, we illustrate the use of FP for solving energy efficiency maximization problems, both for the single-link case which has been treated in prior FP literature, and for the multiple-link case which requires the new techniques developed in this paper.

### A. Single-Link Case

Consider an isolated end-to-end wireless link; the sender and the receiver are equipped with one antenna each. Let $h\in\mathbb{C}$ be the link channel, and let $\sigma^2$ be the AWGN power level. The total power consumption consists of two parts: the transmit power $p$ which is constrained by a power budget $P_{\max}$, and a constant link ON-power $P_{\mathrm{on}}$. The objective is to maximize the ratio of data rate to the total power consumption, namely the energy efficiency, by optimizing $p$, i.e.,

$$\underset{p}{\text{maximize}} \qquad \frac{\log\left(1+|h|^2 p/\sigma^2\right)}{p+P_{\mathrm{on}}} \tag{63a}$$

$$\text{subject to} \qquad 0\le p\le P_{\max}. \tag{63b}$$

This problem is nonconvex in general.

For this single-link case, (63) is a single-ratio concave-convex FP problem and thus its globally optimal solution can be found using the conventional FP technique (e.g., Dinkelbach's transform), as already shown in the past literature [9]–[12]. An alternative is to apply our proposed quadratic transform. The problem is then reformulated as

$$\underset{p,\, y}{\text{maximize}} \quad 2y\sqrt{\log\left(1+\frac{|h|^2 p}{\sigma^2}\right)} - y^2\left(p+P_{\text{on}}\right) \tag{64}$$
$$\text{subject to} \quad 0 \le p \le P_{\text{max}}.$$

Clearly, the optimal $y$ for fixed $p$ is

$$y^\star = \frac{\sqrt{\log\left(1+|h|^2 p/\sigma^2\right)}}{p+P_{\text{on}}}. \tag{65}$$

Then solving $p$ for fixed $y$ is a convex problem. This iteration converges to the global optimum according to Corollary 4.

### B. Multiple-Link Case

Energy efficient maximization across multiple interfering links is a more challenging problem. Consider a spatial multiplex multiple-antenna broadcast channel model with one sender equipped with $M$ antennas to send individual data to its $M$ receivers. Assume that every receiver has $N$ antennas and supports one data stream. Let $\mathbf{H}_m \in \mathbb{C}^{N \times M}$ be the channel between the sender and the $m$th receiver; let $\mathbf{v}_m \in \mathbb{C}^M$ be the beamformer for the transmission to the $m$th receiver. The energy efficiency maximization problem in this case is formulated as

$$\underset{\mathbf{V}}{\text{maximize}} \quad \frac{\sum_{m=1}^{M} R_m(\mathbf{V})}{\sum_{m=1}^{M} \|\mathbf{v}_m\|_2^2 + P_{\text{on}}} \tag{66a}$$

$$\text{subject to} \quad \sum_{m=1}^{M} \|\mathbf{v}_m\|_2^2 \le P_{\text{max}} \tag{66b}$$

where $\mathbf{V}$ refers to the collection $\{\mathbf{v}_m\}$, and the function $R_m(\mathbf{V})$ denoting the data rate of receiver $m$ is

$$R_m(\mathbf{V}) = \log\left(1+\mathbf{v}_m^\dagger \mathbf{H}_m^\dagger \left(\sigma^2\mathbf{I} + \sum_{n\neq m} \mathbf{H}_m \mathbf{v}_n \mathbf{v}_n^\dagger \mathbf{H}_m^\dagger\right)^{-1}\right.$$
$$\left. \cdot\, \mathbf{H}_m \mathbf{v}_m\right). \tag{67}$$

We first describe the approach in [9]–[12]. Dinkelbach's transform recasts the objective function to

$$f_d(\mathbf{V}, y) = \sum_{m=1}^{M} R_m(\mathbf{V}) - y\left(\sum_{m=1}^{M} \|\mathbf{v}_m\|_2^2 + P_{\text{on}}\right). \tag{68}$$

However, unlike the single-link case, the reformulation $f_d$ is no longer a concave function of $\mathbf{V}$, so optimizing $\mathbf{V}$ for fixed $y$ is numerically difficult. Hence, the iterative algorithm based on Dinkelbach's transform cannot be easily extended to the multiple-link scenario. In fact, [12] considers multiple links only under the assumption that the resulting SINRs are all sufficiently high; [11] globally solves the $f_d$ maximization problem using a monotonic optimization approach (which has an exponential-time complexity), and also proposes a polynomial-time algorithm to attain a stationary point when the transmitter has a single antenna (i.e., when $\mathbf{v}_m$ reduces to a scalar). Moreover, [35] proposes a gradient method to maximize the nonconcave function $f_d$ in (68), and [36] advocates successive convex approximation. But none of them can find in polynomial time the globally optimal $\mathbf{V}$ that maximizes $f_d$. We remark that the optimality of $\mathbf{V}$ in maximizing $f_d$ is critical to the convergence of the Dinkelbach's algorithm [22], so these existing polynomial-time algorithms are not guaranteed to converge in general. By contrast, our approach does not rely on the Dinkelbach's transform, and has provable convergence. As a further remark, if the sum rate objective function is changed to the superposition coding inner bound, the new problem after the Dinkelbach's transform would have been convex and can be optimally solved by a water-filling scheme [37].

The paper advocates a novel use of the quadratic transform to address the problem. First, apply the single-ratio quadratic transform (i.e., Theorem 1) to decouple the energy efficiency as

$$f_q(\mathbf{V}, y) = 2y\left(\sum_{m=1}^{M} R_m(\mathbf{V})\right)^{\frac{1}{2}} - y^2\left(\sum_{m=1}^{M} \|\mathbf{v}_m\|_2^2 + P_{\text{on}}\right). \tag{69}$$

The same issue as with the Dinkelbach's transform approach now arises: the reformulated objective function is not concave over $\mathbf{v}_m$. It is crucial to observe that the function $x^{\frac{1}{2}}$ is nondecreasing and concave, and also that the second term in (69) is concave. Thus, the concavity of $f_q$ over $\mathbf{v}_m$ can be restored if the term inside the square root $\sum_{m=1}^{M} R_m$ is recast as a concave function.

Following this idea, we apply the (multidimensional) quadratic transform to each SINR term inside the $R_m$ expression (67) in $f_q$, and further recast $f_q$ to $f_{qq}$ as in (70) at the bottom of the page. The ultimate reformulation of (66) after the two uses of the quadratic transform now becomes

$$\underset{\mathbf{V},\, y,\, \mathbf{Z}}{\text{maximize}} \quad f_{qq}(\mathbf{V}, y, \mathbf{Z}) \tag{71a}$$

$$\text{subject to} \quad \sum_{m=1}^{M} \|\mathbf{v}_m\|_2^2 \le P_{\text{max}} \tag{71b}$$

$$\mathbf{z}_m \in \mathbb{C}^N \tag{71c}$$

where $\mathbf{Z}$ refers to the collection $\{\mathbf{z}_m\}$. We remark that $y$ and $\mathbf{z}_m$ are the auxiliary variables introduced by the first and the second use of FP, respectively.

$$f_{qq}(\mathbf{V}, y, \mathbf{Z}) = 2y\left(\sum_{m=1}^{M} \log\left(1+2\text{Re}\{\mathbf{z}_m^\dagger \mathbf{H}_m \mathbf{v}_m\} - \mathbf{z}_m^\dagger\left(\sigma^2\mathbf{I} + \sum_{n\neq m} \mathbf{H}_m \mathbf{v}_n \mathbf{v}_n^\dagger \mathbf{H}_m^\dagger\right)\mathbf{z}_m\right)\right)^{\frac{1}{2}} - y^2\left(\sum_{m=1}^{M} \|\mathbf{v}_m\|_2^2 + P_{\text{on}}\right) \tag{70}$$

---

**Algorithm 6: Nested FP for Energy Efficiency Maximization**

---

**Step 0:** Initialize $\mathbf{V}$ to a feasible value.

**Repeat**

    **Step 1:** Update $\mathbf{Z}$ by (72).

    **Step 2:** Update $y$ by (73).

    **Step 3:** Update $\mathbf{V}$ by solving the convex optimization problem (71) over $\mathbf{V}$ for fixed $\mathbf{Z}$ and $y$.

**until** the value of function $f_{qq}$ in (70) converges.

---

We propose an iterative optimization. When all the other variables are held fixed, the optimal $\mathbf{z}_m$ is

$$\mathbf{z}_m^\star = \left(\sigma^2 \mathbf{I} + \sum_{n \neq m} \mathbf{H}_m \mathbf{v}_n \mathbf{v}_n^\dagger \mathbf{H}_m^\dagger\right)^{-1} \mathbf{H}_m \mathbf{v}_m, \ \forall m. \quad (72)$$

After the update of $\mathbf{z}_m$, the optimal $y$ is

$$y^\star = \frac{\sqrt{\sum_{m=1}^M R_m(\mathbf{V})}}{\sum_{m=1}^M \|\mathbf{v}_m\|_2^2 + P_{\text{on}}}. \quad (73)$$

Most importantly, when $\mathbf{z}$ and $y$ are both fixed, (71) is a convex problem of $\mathbf{v}_m$, and therefore the optimal $\mathbf{v}_m$ can be efficiently found using the standard numerical method.

This iterative optimization is summarized in Algorithm 6. We refer to it as the nested FP approach, because the reformulating procedure involves an outer FP for the energy efficiency ratio as well as an inner FP for the nesting SINR terms. Based on the equivalence of objective function property C3 in Section II-B, it is easy to verify the convergence of Algorithm 6 to a stationary point of the original problem (66) with the energy efficiency value nondecreasing after each iteration.

### C. Numerical Example

The simulation model assumes flat-fading channel(s) over a 1MHz-wide frequency band. The maximum transmit power level is set to be 21dBm; the on-power level is set to be 5dBm; the background noise level is set to be −100dBm. We test the proposed algorithm for two network scenarios:

- Single-link case: Consider one pair of sender and receiver, equipped with one antenna each; the channel coefficient between them is modeled with −120dB pathloss.
- Multiple-link case: Consider 1 sender and 3 receivers; the sender has 3 antennas and the receivers have 2 antennas each. The channel coefficients between the transmit and receive antennas are modeled with i.i.d. Rayleigh fading component plus −120dB pathloss.

Fig. 6 compares the Dinkelbach's transform approach [9]–[12] and the proposed quadratic transform in maximizing energy efficient for the single-link case. It can be observed that Dinkelbach's transform gives a faster convergence. To attain the optimal energy efficiency, Dinkelbach's transform needs 4 iterations while the quadratic transform needs 8 iterations.

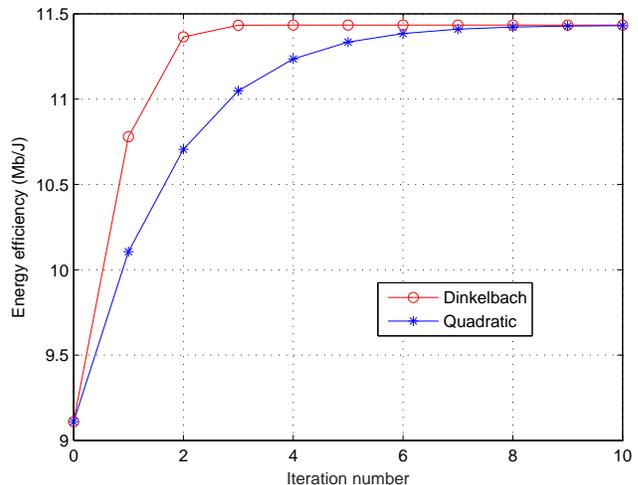

Fig. 6: Energy efficiency maximization for a single link.

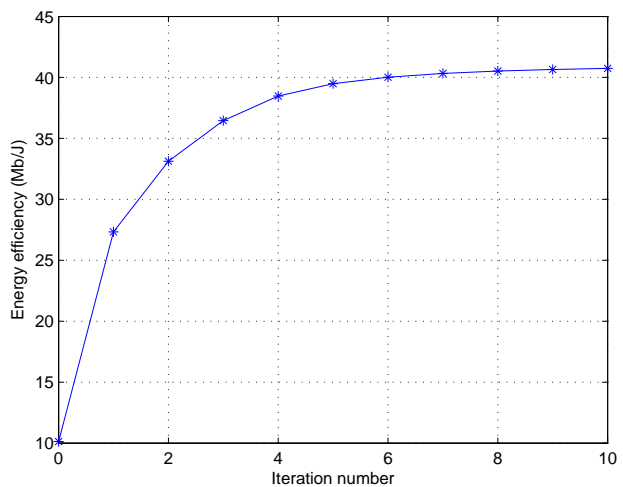

Fig. 7: Energy efficiency maximization for a broadcast network.

This result agrees with the convergence rate analysis in Section II-E.

Fig. 7 evaluates the performance of Algorithm 6 in maximizing the multiple-link energy efficiency. We reiterate that Dinkelbach's transform [9]–[12] is not applicable in this case. As can be seen from the figure, Algorithm 6 raises the energy efficiency significantly to more than four-fold after just 8 iterations.

## VI. CONCLUSION

The paper introduces a novel FP technique called quadratic transform, which can tackle a broad range of FP problems with multiple ratios in contrast to the conventional techniques which can only handle single ratio or the max-min case. Based on the quadratic transform, a variety of FP approaches are devised for solving the continuous problems in communication systems, i.e., power control, beamforming, and energy efficiency maximization. The proposed FP approaches recast the original nonconvex problem to a sequence of convex problems, thereby allowing efficient iterative optimization with provable convergence to a stationary point solution. Part I of this paper treats continuous optimization problems. Discrete problems are treated in Part II [19].

# APPENDIX A
## PROOF OF THEOREM 1

It is easy to verify that $g(\mathbf{x}, y)$ in (7) satisfies C1-C4. Below we focus on showing that the form of $g(\mathbf{x}, y)$ in (8) is necessary and sufficient when C4 is strengthened to require that $\partial^2 g/\partial y^2$ is independent of $y$. First, under the strengthened C4 and by C1, function $g$ must be of the form:

$$g(\mathbf{x}, y) = f(A(\mathbf{x}))(\alpha_2 y^2 + \alpha_1 y + \alpha_0) \\ + h(B(\mathbf{x}))(\beta_2 y^2 + \beta_1 y + \beta_0) \quad (74)$$

for some parameters $\alpha_i$ and $\beta_i$ such that

$$\frac{\partial^2 g(\mathbf{x}, y)}{\partial y^2} = 2\alpha_2 f(A(\mathbf{x})) + 2\beta_2 h(B(\mathbf{x})) \le 0. \quad (75)$$

For ease of notation, we omit the function arguments of $A(\mathbf{x})$ and $B(\mathbf{x})$ in the rest of the proof. First, note that $\partial^2 g(\mathbf{x}, y)/\partial y^2$ cannot be zero, as otherwise $\max_y g(\mathbf{x}, y) = \infty$ and thus C3 cannot be satisfied. Given a particular $\mathbf{x}$, the maximum value of $g(\mathbf{x}, y)$ over $y$ can now be obtained in closed form as

$$\max_y g(\mathbf{x}, y) = \alpha_0 f(A) + \beta_0 h(B) - \frac{(\alpha_1 f(A) + \beta_1 h(B))^2}{4(\alpha_2 f(A) + \beta_2 h(B))}. \quad (76)$$

As required by C3, we must have $\max_y g(\mathbf{x}, y) = A/B$. One way to satisfy this relation is to have $\alpha_0 = 0, \beta_0 = 0, \alpha_1 = 2, \beta_1 = 0, \alpha_2 = 0, \beta_2 = 1$, $f(A) = \sqrt{A}$, and $h(B) = B$. This gives the proposed quadratic transform (7). The remainder of the proof aims to show that a more general form of this solution (8) is the unique solution satisfying the above.

The main idea is to determine functions $f$ and $h$ as well as parameters $\alpha_i$ and $\beta_i$ by substituting different $(A, B)$ pairs in (76). First, put $A = 0$ (i.e., $A(\mathbf{x})$ is a zero constant function) then $\max_y g = A/B = 0$ for any $B$, i.e.,

$$(4\beta_0 \beta_2 - \beta_1^2) h^2(B) + (4\alpha_2 \beta_0 + 4\alpha_0 \beta_2 - 2\alpha_1 \beta_1) f(0) h(B) \\ + (4\alpha_0 \alpha_2 f^2(0) - \alpha_1^2 f^2(0)) = 0. \quad (77)$$

For this to hold for any $B$, we must have

$$4\beta_0 \beta_2 - \beta_1^2 = 0. \quad (78)$$

In this case, the expression (76) reduces to

$$\max_y g(\mathbf{x}, y) = \frac{C}{D} \quad (79)$$

where

$$C = (4\alpha_0 \alpha_2 - \alpha_1^2) f^2(A) + (4\alpha_0 \beta_2 + 4\alpha_2 \beta_0 - 2\alpha_1 \beta_1) f(A) h(B) \quad (80)$$

and

$$D = 4(\alpha_2 f(A) + \beta_2 h(B)). \quad (81)$$

Second, consider the case that $B \to 0_+$, then $\max_y g(\mathbf{x}, y) = A/B = \infty$ for any $A \ne 0$. For this to happen, we need $D \to 0$ for any $A$, whenever $B \to 0_+$. This means that the first term in $D$, which is a function of $A$ only, must be zero, or

$$\alpha_2 = 0. \quad (82)$$

Third, consider the case that $A \to 0_+$, then $\max_y g(\mathbf{x}, y) = A/B = 0$ for any $B$. For this to happen, we need $C \to 0$ for any $B$, whenever $A \to 0_+$. This means that the second term in $C$, which is a function of $B$ must be zero. Since $f(A)$ cannot be a constant zero, we must have

$$4\alpha_0 \beta_2 + 4\alpha_2 \beta_0 - 2\alpha_1 \beta_1 = 4\alpha_0 \beta_2 - 2\alpha_1 \beta_1 = 0. \quad (83)$$

The $\max_y g(\mathbf{x}, y)$ expression now becomes

$$\max_y g(\mathbf{x}, y) = \frac{-\alpha_1^2 f^2(A)}{4\beta_2 h(B)}. \quad (84)$$

It can be readily seen that for it to be equal to $A/B$, we must have

$$f(A) = s_1 \sqrt{A} \quad \text{and} \quad h(B) = s_2 B \quad (85)$$

for some nonzero $s_1, s_2$ such that

$$-\alpha_1^2 s_1^2 = 4\beta_2 s_2. \quad (86)$$

Summarizing, $g(\mathbf{x}, y)$ must have this form:

$$g(\mathbf{x}, y) = s_1(\alpha_1 y + \alpha_0)\sqrt{A(\mathbf{x})} + s_2(\beta_2 y^2 + \beta_1 y + \beta_0)B(\mathbf{x}) \quad (87)$$

subject to (78), (83) and (86). Using (78), (83) and (86), i.e.,

$$\begin{cases} \beta_1^2 = 4\beta_0 \beta_2 \\ 2\alpha_0 \beta_2 = \alpha_1 \beta_1 \\ -\alpha_1^2 s_1^2 = 4\beta_2 s_2 \end{cases}, \quad (88)$$

we obtain

$$\beta_2 = -\frac{\alpha_1^2 s_1^2}{4s_2}, \quad \beta_1 = -\frac{\alpha_1 \alpha_0 s_1^2}{2s_2}, \quad \beta_0 = -\frac{\alpha_0^2 s_1^2}{4s_2}. \quad (89)$$

With the above identities substituted in (74) to get rid of $\beta_i$'s, the reformulation $g(\mathbf{x}, y)$ becomes

$$g(\mathbf{x}, y) = s_1(\alpha_1 y + \alpha_0)\sqrt{A(\mathbf{x})} - \frac{s_1^2(\alpha_1 y + \alpha_0)^2}{4} B(\mathbf{x}). \quad (90)$$

The above form of $g(\mathbf{x}, y)$ can be rewritten as (8) by defining two new parameters: $t_1 = s_1\alpha_1/2$ and $t_2 = s_1\alpha_0/2$. Finally, we note that $g(\mathbf{x}, y)$ in (8) satisfies the strengthened C1-C4 when $t_1 \ne 0$. This form of $g(\mathbf{x}, y)$ is therefore necessary and sufficient for this set of conditions.

**Kaiming Shen** (S'13) received the B.Eng. degree in information security and the B.S. degree in mathematics both from Shanghai Jiao Tong University, Shanghai, China in 2011, and the M.A.Sc. degree in electrical and computer engineering from the University of Toronto, Ontario, Canada in 2013. He is currently pursuing the Ph.D. degree at the University of Toronto.

His research interests include mathematical optimization, algorithms for wireless networks, and information theory.

**Wei Yu** (S'97-M'02-SM'08-F14) received the B.A.Sc. degree in Computer Engineering and Mathematics from the University of Waterloo, Waterloo, Ontario, Canada in 1997 and M.S. and Ph.D. degrees in Electrical Engineering from Stanford University, Stanford, CA, in 1998 and 2002, respectively. Since 2002, he has been with the Electrical and Computer Engineering Department at the University of Toronto, Toronto, Ontario, Canada, where he is now Professor and holds a Canada Research Chair (Tier 1) in Information Theory and Wireless Communications. His main research interests include information theory, optimization, wireless communications and broadband access networks.

Prof. Wei Yu currently serves on the IEEE Information Theory Society Board of Governors (2015-20). He was an IEEE Communications Society Distinguished Lecturer (2015-16). He is currently an Area Editor for the IEEE Transactions on Wireless Communications (2017-20). He served as an Associate Editor for IEEE Transactions on Information Theory (2010-2013), as an Editor for IEEE Transactions on Communications (2009-2011), and as an Editor for IEEE Transactions on Wireless Communications (2004-2007). He is currently the Chair of the Signal Processing for Communications and Networking Technical Committee of the IEEE Signal Processing Society (2017-18) and served as a member in 2008-2013. Prof. Wei Yu received the Steacie Memorial Fellowship in 2015, the IEEE Signal Processing Society Best Paper Award in 2017 and 2008, a Journal of Communications and Networks Best Paper Award in 2017, an IEEE Communications Society Best Tutorial Paper Award in 2015, an IEEE ICC Best Paper Award in 2013, the McCharles Prize for Early Career Research Distinction in 2008, the Early Career Teaching Award from the Faculty of Applied Science and Engineering, University of Toronto in 2007, and an Early Researcher Award from Ontario in 2006. Prof. Wei Yu is a Fellow of the Canadian Academy of Engineering, and a member of the College of New Scholars, Artists and Scientists of the Royal Society of Canada. He is recognized as a Highly Cited Researcher.